# Phonons and Stability of Infinite-Layer Iron Oxides $SrFeO_2$ and $CaFeO_2$

M. K. Gupta, R. Mittal and S. L. Chaplot
Solid State Physics Division, Bhabha Atomic Research Centre, Trombay, Mumbai 400085
Cédric Tassel and Hiroshi Kageyama
Department of Energy and Hydrocarbon Chemistry, Graduate School of Engineering, Kyoto University, Kyoto, 615-8510, Japan

We present detailed ab-initio lattice dynamical analysis of the Fe-O infinite-layer compounds $CaFeO_2$ and $SrFeO_2$ in various magnetic configurations. These indicate strong spin-phonon coupling in $SrFeO_2$ in contrast to that in case of $CaFeO_2$. Powder neutron inelastic scattering experiments on $SrFeO_2$ have also been performed at temperatures from 5 K to 353 K in the antiferromagnetic phase and analyzed using the ab-initio calculations. These suggest distortion of the ideal infinite planer structure above 300 K. From our ab-initio calculations in $SrFeO_2$ as a function of volume, we suggest that the distortion in $SrFeO_2$ above 300 K is similar to that known in $CaFeO_2$ at ambient conditions. The distortion of the planer structure of $CaFeO_2$ involves doubling of the planer unit cell that may be usually expected to be due to a soft phonon mode at the M-point (1/2 1/2 0). However, our ab-initio calculations show quite unusually that all the M-point (1/2 1/2 0) phonons are stable, but two stable $M_3^+$ and $M_2^-$ modes anharmonically couple with an unstable $B_u$ mode at the zone centre and lead to the cell doubling and the distorted structure. Magnetic exchange interactions in both the compounds have been computed on the basis of the ideal planar structure (*P*4/*mmm* space group) and with increasing amplitude of the $B_u$ phonon mode. These reveal that the magnetic exchange interactions reduce significantly with increasing distortion. We have extended the ab-initio phonon calculation to high pressures, which reveal that, above 20 GPa of pressure, the undistorted planer $CaFeO_2$ becomes dynamically stable. We also report computed phonon spectra in $SrFeO_3$ that has a cubic structure, which is useful to understand the role of the difference in geometry of oxygen atoms around the Fe atom with respect to planer $SrFeO_2$.

## I. INTRODUCTION

Low-dimensional magnetic systems have received much attention due to their exotic magnetic and electronic properties. There is a plenty of room to explore unconventional properties as represented by high-temperature superconductivity through electron/hole doping in the parent antiferromagnetic (AFM) square lattice of $La_2CuO_4$[1] and $K_xFe_{2-y}Se_2$ [2]. However, the mechanism of superconductivity in cuprates and Fe-based compounds is still under intensive debate [3].

Iron forms a large number of oxides mostly with $FeO_4$ tetrahedral, $FeO_5$ pyramidal or $FeO_6$ octahedral configurations. For example, $AFeO_y$ (A = Sr, Ca; $y$ ~ 2.5) adapts a brownmillerite structure consisting of tetrahedral and octahedral layers [4]. Earlier, the gillespite mineral $BaFeSi_4O_{10}$ was the only example with iron in square planar coordination [5], which is stabilized by four-member rings of $SiO_4$. Later the synthesis of a metastable phase $SrFeO_2$ using a topochemical reaction of $SrFeO_y$ was reported[6]. $SrFeO_2$ is distinct (from $BaFeSi_4O_{10}$) in that square-planar $FeO_4$ units are connected with each other to form extended $FeO_2$ layers that are separated by strontium atoms (Fig. 1). The resultant structure is isostructural with the infinite-layer structure $SrCuO_2$ ($P4/mmm$).The $Fe^{2+}$ ion is in a high spin state ($S$ = 2) with the electronic configurations of $(d_{z2})^2(d_{xz},d_{yz})^2(d_{xy})^1(d_{x2-y2})^1$ [7]. $SrFeO_2$ is an AFM insulator with a high ordering temperature $T_N$ of 473 K, while at high pressure it undergoes a spin transition to $S$ = 1 accompanied by a transition to a ferromagnetic (FM) half-metallic state[8]. Magnetic properties of $SrFeO_2$ have been examined by density functional theory (DFT) band structure and total energy calculations[7, 9]. Recently high pressure study[10] on $SrFeO_2$ based on first principles DFT simulation is performed to explain the antiferromagnetic to ferromagnetic phase transition at high pressure. In the last decade plenty of studies[10-19] have been reported on planer $AFeO_2$ (A = Ca, Sr) and their derivatives.

When Sr is replaced by Ca with a smaller ionic radius, the infinite-layer structure becomes corrugated[20]. In $CaFeO_2$ ($P$-$42_1m$), oxygen atoms move along the $z$ direction to distort $FeO_4$ square planar unit toward a tetrahedral shape. This distortion affects the exchange interaction and leads to a reduction in $T_N$ (420 K). The origin of this distortion in $CaFeO_2$ is discussed in terms of phonon. Ab-initio density functional perturbation theory (DFPT) calculation of the zone centre phonon modes has also been investigated[9, 20]. For $CaFeO_2$ assuming the $P4/mmm$ space group, two unstable phonon modes are indicated, one of which involves out-of-plane translation motion of the oxygen atoms along z-axis, while the other zone boundary mode shows in-plane rotation of the $FeO_4$ squares. Recently, a high-resolution neutron diffraction study at various temperatures[21] has demonstrated that even in $SrFeO_2$ the ideal infinite-layer structure is destabilized upon approaching to the Neel temperature (473 K). The analysis shows a local transverse mode creates buckling in the $FeO_4$ planes, resulting in lowering the tetragonal symmetry. Such transverse distortion created by local structural instability significantly weakens the exchange interactions.

In order to obtain further insight into structural instability of $SrFeO_2$ and $CaFeO_2$ as well as its effect on exchange parameters, we have performed the *ab-initio* phonon calculations for planar $SrFeO_2$ (*P*4/*mmm*) and both planar and distorted $CaFeO_2$ (*P*4/*mmm* and $P$-42$_1m$), termed here after by p-$CaFeO_2$ and d-$CaFeO_2$, respectively, in various magnetic configurations in the entire Brillouin zone. The longitudinal and transverse optic (LO-TO) splitting has been taken into account in the calculations of phonon frequencies. These calculations are useful to interpret the measured spectrum and able to explain the origin of distortion in $CaFeO_2$ as well as high-pressure stability of $CaFeO_2$. Our calculations show that $SrFeO_2$ (*P*4/*mmm*) and d-$CaFeO_2$ ($P$-42$_1m$) are dynamically stable with the G-type AFM structure, while p-$CaFeO_2$ (*P*4/*mmm*) is dynamically unstable at ambient pressure. The calculated phonon density of states of $SrFeO_2$ has been compared with the powder inelastic neutron scattering result.

Here we have also performed the lattice dynamical calculation in $SrFeO_3$ system and compared it with the $SrFeO_2$ spectra. The calculations highlight the role of difference in the geometry of oxygen atoms around Fe atom in both the compounds, which leads to the difference in the phonon spectra in these compounds. The paper is organised as follows. In section II and III, we describe experimental and calculation methods. In section IVA, we discuss the result of neutron inelastic scattering experiments of $SrFeO_2$. Section IVB gives detailed lattice dynamical calculations of $CaFeO_2$ and $SrFeO_2$. In Section IVC, dynamical instabilities in planer $CaFeO_2$ and stabilization of distorted $CaFeO_2$ at ambient pressure has been discussed. Phase transition from distorted to planer $CaFeO_2$ at high pressure is discussed in Section IVD. The work on Spin phonon coupling and magnetic exchange interaction Parameters in planar $SrFeO_2$ and $CaFeO_2$ is given in Section IVE. In section IVF, we discuss the comparison of the phonon spectra of $SrFeO_2$ and $SrFeO_3$ followed by conclusions in Section V.

## II. EXPERIMENTAL

The polycrystalline sample of $SrFeO_2$ was prepared by a topochemical reduction of $SrFeO_{2.875}$ with $CaH_2$(2 molar excess) at 300 °C for 48 hours in an evacuated tube[6]. After this reaction, the residual $CaH_2$ and the byproduct were eliminated by methanol. The inelastic neutron scattering experiments were performed using the MARI time of flight spectrometers at ISIS, UK. The measurements were done in the neutron-energy-loss mode using incident neutron energy of 120 meV at several temperatures from 5 to 353 K. The incoherent approximation [22-24] was used for extracting neutron weighted phonon density of states from the measured scattering function *S*(*Q*,*E*).

## III. COMPUTATIONAL DETAILS

The Vienna *ab initio* simulation package (VASP) [25, 26] was used for calculations. The plane wave pseudo-potential with plane wave kinetic energy cutoff of 400 eV was used for both compounds. The integrations over the Brillouin zone were sampled on a 4×4×4 grid of k-points generated by Monkhorst-pack method [27]. The generalized gradient approximation (GGA) exchange correlation given by Perdew, Becke and Ernzerhof [28, 29] with projected-augmented wave method has been used. Since the compounds contain localized Fe 3*d* electrons at ambient pressure, we have used the simplified approach to the LSDA+U, introduced by Dudarev *et al*. [30]. The onsite interaction term U was taken to be 4.0 eV for *d* electrons of iron.

The primitive unit cell for the ideal infinite-layer structure (*P*4/*mmm*) has 4 atoms. The magnetic structure[9] of $SrFeO_2$ has a super-cell of√2×√2×2 with respect to its primitive unit cell, thus containing16 atoms in the magnetic unit cell. For the phonon calculation, we have used a 2√2×2√2×4 super-cell with 128 atoms. In case of $CaFeO_2$, the experimentally obtained distorted infinite-layer structured-$CaFeO_2$($P\text{-}42_1m$) has a √2×√2×1 unit cell (8 atoms), while the magnetic cell has a 1×1×2 super-cell (thus 16 atoms) with respect to its crystal cell of d-$CaFeO_2$. A 2×2×4 super-cell has (with respect to d-$CaFeO_2$) been used for the phonon calculation of $CaFeO_2$ (128 atoms). While calculating inter-atomic force constants, we have included the G-type AFM magnetic ordering in both compounds. The crystal structure of both the compounds as used in the calculations is shown in Fig.1. Total energies and inter-atomic forces were calculated for 14 (*P*4/*mmm*) and 12 ($P\text{-}42_1m$) distinct structures resulting from displacements of the symmetrically inequivalent atoms along the three Cartesian directions (±*x*, ±*y* and ±*z*).

Further, total energy of the cubic perovskite $SrFeO_3$ (*Pm*-3*m*) has been calculated in various antiferromagnetic (A, C and G) as well as in FM phase at *T*=0 K. The Hellman-Feynman forces have been calculated for eight different displacement configurations. The phonon spectra of $SrFeO_3$ have been calculated in computationally stable FM configuration of $SrFeO_3$.The convergence criteria for the total energy and ionic forceswere set to $10^{-8}$ eV and $10^{-5}$ eV $Å^{-1}$, respectively.

Phonon spectra for $SrFeO_2$ and $CaFeO_2$ were extracted using the PHONON software[31]. All the phonon calculations are carried out in the fully relaxed configuration. The relaxed unit cell parameters are given in TABLE I. We have also carried out an additional calculation of phonon modes in $SrFeO_2$ and p-$CaFeO_2$ (*P*4/*mmm*) as described in Section IV. The exchange parameter has been calculated using a Heisenberg spin Hamiltonian with forth neighbour interaction. To determine the magnetic ground state and discuss the magnetic properties of $SrFeO_2$, we considered four more ordered spin structures besides the FM state, namely, the AF1 state with *q*= (½ ½ ½), the AF2 state with *q* = (0 0 ½), the AF3 state with *q* = (½ ½ 0) and the AF4 state with *q* = (½ 0 ½).

## IV. RESULTS AND DISCUSSION

### A. Experimental and Calculated Phonon Spectra of planar $SrFeO_2$

We have measured the inelastic spectra of $SrFeO_2$ over a temperature range from 5K to 353 K. The measured neutron inelastic spectra are shown in Fig.2. $SrFeO_2$ exhibits the G-type AFM order below 473 K.The data were averaged over a high-*Q* regime of 9-10$Å^{-1}$ to avoid paramagnetic contributions to the phonon data. In Fig 3, we have shown the computed neutron weighted phonon spectra together with experimental spectra at *T*=5K. Here, the measured spectra were corrected from multiphonon contributions as calculated using Sjolander formalism[32]. In order to probe the individual atomic contribution, we also calculated the neutron weighted partial density of states. The calculations were done under GGA approximation. The GGA is expected to overestimate the lattice parameters and in turn underestimate the phonon frequencies. When the calculated energy spectrum is scaled by 6%, we obtained an excellent agreement between theory and experiment (Fig. 3), which partially justifies the use of our theoretical tool for other thermodynamical properties.

We have also calculated the partial phonon density of states associated with various atoms (in Figure 4). This helps to assign the peaks in the experimental spectra. We note that the first peak around 15 meV (Fig. 4) has contributions from Sr atoms. The peaks around 25 meV, 30 meV and 40meV have large contributions from oxygen and iron atoms. The phonon spectra above 40 meV have mainly contributions from oxygen atoms. At higher temperature (Fig.2), one can observe that the peaks about 15 meV and 25 meV retain their spectral features, while the peak about 30 meV shifts to higher energy.

Furthermore, the peaks are significantly broadened above 300 K, which might be related[21] to the distortion in the planer structure of $SrFeO_2$ The nature of the distortion is discussed latter on the basis of ab-initio phonon calculations in Section IV-C.

**B. Phonon Spectra in $SrFeO_2$ and $CaFeO_2$**

We have calculated the phonon spectra of $SrFeO_2$ and d-$CaFeO_2$ and p-$CaFeO_2$ in a relaxed geometry. The calculated lattice constants and experimental values are given in Table I. In Fig. 4, we show the calculated partial and total density of states for $CaFeO_2$ and $SrFeO_2$. We find that the p-$CaFeO_2$ is dynamically unstable as observed in the previous study[20]. It may be noted that Ref.[20] provides the calculations of phonon modes at few selected points in the Brillouin zone. However our calculations of phonon spectra in entire Brillouin zone successfully give a complete picture of dynamics. They allow one to understand the differences in the calculated phonon spectra of d-$CaFeO_2$ and p-$CaFeO_2$ and compare with those of $SrFeO_2$.

The calculated phonon spectra of d-$CaFeO_2$ and $SrFeO_2$ are significantly different (left panel Fig.4). In $SrFeO_2$, the contribution from oxygen is extended in the entire spectral range up to 80 meV, while in d-$CaFeO_2$ it ranges up to 70 meV. Given the larger unit cell of $SrFeO_2$, one would naively expect that the phonon spectra in d-$CaFeO_2$ are broader than those in$SrFeO_2$, which is in contrast to the calculated result. This difference is understood in terms of the difference in bonding nature in Fe-O: the Fe-O stretching bond is stronger in planer geometry than that of distorted one. The A-site contribution in $SrFeO_2$ is limited to 35 meV, while in $CaFeO_2$ this extends up to 45 meV, which seems to follow the mass difference of Ca (40.08 amu) and Sr(87.62 amu). No major difference is observed in the Fe spectra.

A further comparison of calculated phonon spectra in p-$CaFeO_2$ and $SrFeO_2$, as shown in the right panel of Fig.4, reveals that it simply follows the volume consideration. In the p-$CaFeO_2$, the oxygen partial density of states extends from 10*i* to 90 meV, while the unstable modes in p-$CaFeO_2$ are found to be stable in the d-$CaFeO_2$. It seems that oxygen dynamics plays an important role in stabilizing the distorted structure.

The partial densities of states have been used for the calculation of the mean square amplitude for various atoms at different temperatures. The comparison between the calculated values and the experimental data at 300 K is given in Table I. The calculated temperature factors for Sr/Ca, Fe and O are similar in both compounds. The difference in nature of the in-plane and out-of-plane bonding may result in large anisotropic values of $u^2$ along the *x* and *z* axis. The temperature variation of the calculated anisotropic mean squared amplitude of various atoms in $SrFeO_2$ and d-$CaFeO_2$ is shown in Fig. 5. The oxygen atoms in d-$CaFeO_2$ show large anisotropic behavior. The anisotropy is least in Ca. In case of $SrFeO_2$, however, we found that oxygen atoms show maximum amplitude of vibrations along z direction and show large anisotropic behavior. The values of $u^2$for Sr, Fe and O are lower than those of Ca, Fe and O atoms in d-$CaFeO_2$. The observation is in agreement with the recent measurements[21] of $u^2$ in $SrFeO_2$.The larger values of $u^2$ of various atoms in d-$CaFeO_2$ in comparison with$SrFeO_2$may be due to smaller unit cell of d-$CaFeO_2$in addition to the difference in bonding nature.

In order to compare the nature of bonding in d-$CaFeO_2$, p-$CaFeO_2$ and $SrFeO_2$, we have computed the Born dynamical charge tensor (Table III). Thanks to the high local site symmetry of Ca/Sr, Fe and O atoms, the Born effective charge tensors in planar structure (*P*4/*mmm*) have only few non-zero diagonal elements, namely in-plane ($Z_{xx} = Z_{yy}$) and out-of-plane ($Z_{zz}$) components. The calculated components of Born effective charges for p-$CaFeO_2$ and $SrFeO_2$ are very close. However, we observed a large difference in the Fe charges along *z* direction ($Z_{xx}$ component) between p-$CaFeO_2$ and $SrFeO_2$. This might be due to difference in inter-planar separation of $FeO_4$ layers in p-$CaFeO_2$ and $SrFeO_2$ along the *c* direction, attributed to the difference in ionic radii between Ca and Sr. Furthermore, a comparison of the Born effective charge tensors in p-and d-$CaFeO_2$ reveals that Born effective charges for Ca and Fe are nearly the same. The out-of-plane ($Z_{zz}$) component for O atoms is nearly same, while there is a large difference in the in-plane ($Z_{xx}$, $Z_{yy}$) components. This indicates that the nature of Fe-O bonding in the *x*-*y* plane in planer and distorted structure is quite different.

The p-$CaFeO_2$ and $SrFeO_2$ (*P*4/*mmm*) has 4 atoms in unit cell, which results in 12 vibrational modes for each wave vector. Group theoretical symmetry analysis[31] was carried out to classify the phonon modes belonging to various representations. The group theoretical decomposition of the phonon modes at the zone centre (Γ point) and zone boundary (M and A points) are given by:

$$\Gamma=3A_{2u} + B_u + 4E_u$$

$$M= M_1^+ + M_2^+ + M_3^+ + M_4^+ + 2M_5^+ + M_2^- + M_3^- + 4M_5^-$$

$$A= A_1^+ + A_2^+ + A_3^+ + 2A_4^+ + 4A_5^+ + A_3^- + 2A_5^-$$

At the Γ point, all the modes are infrared active. For d-$CaFeO_2$ ($P$-$42_1m$), the classification of modes at the Γ point is given by

$$\Gamma=3A_1 + A_2 + 2B_1 + 4B_2 + 7E$$

In this case, all the modes are Raman active, with the $B_2$ and E modes being also infra-red active. The calculated zone centre modes are given in Table II(a). The lowest energy $B_u$ mode (9.43 meV) in $SrFeO_2$ is smaller than the lowest $A_1$ mode (15.16 meV) in $CaFeO_2$, which should arises from the larger unit cell volume in $SrFeO_2$ in comparison with $CaFeO_2$. The zone boundary modes for p-$CaFeO_2$ (*P*4/*mmm*) have also been calculated (Table II(b)).

**C. Dynamical Instabilities in Planer $CaFeO_2$ and Stabilization of Distorted $CaFeO_2$ at Ambient Pressure**

The calculated phonon dispersion relations along various high symmetry directions in $SrFeO_2$, p- and d-CaFeO2 are shown in Fig. 6. All the modes in the entire Brillouin zone are found to be stable in $SrFeO_2$ and d-$CaFeO_2$, while in p-$CaFeO_2$ the low energy $B_u$ mode at the zone centre and $A_3^+$ mode at A(½½½) point are unstable. We have performed the amplitude mode analysis[33], which indicates that the distortion in p-$CaFeO_2$ is induced by $B_u$, $M_3^+$ and $M_2^-$ phonon modes. Our calculation of phonon dispersion relation in high symmetry phase of p-$CaFeO_2$ gives stable phonons at M(½½ 0) point. The $M_3^+$ mode involves the in-phase rotation of $FeO_4$ units about *z* axis, while the $M_2^-$ mode involves the out-of-phase displacement of neighbouring Ca atoms along *z*. The unstable $B_u$ mode at zone centre involves the displacement of oxygen atom along ±z direction and another unstable $A_3^+$ mode at A(½ ½ ½) involves the out-of-phase rotation of $FeO_4$ about *z* direction in the alternative layer. In summary, $M_3^+$, $B_u$ and $A_3^+$ modes could be responsible for the shifting of oxygen positions, while $M_2^-$ mode is responsible for the shifting of Ca position. The strong coupling between the low energy phonon modes in p-$CaFeO_2$ has been also discussed in Ref.[20]. These characteristic phonon modes are represented in Fig. 7.

In order to obtain further insights into the structural instability in p-$CaFeO_2$, we have calculated the energy profiles (Fig. 8) of p-$CaFeO_2$ by exciting pairs of phonon modes simultaneously with different amplitude. The unstable modes $A_3^+$ and $B_u$ will not give a minimum in the total energy at zero phonon distortion. One may expect some minima at finite distortion created by these unstable phonon modes. The calculated energy profile of $B_u$ phonon mode with different distortion amplitude of $A_3^+$ phonon mode is shown in Fig. 8(a). As we increase the amplitude of any one of the mode, the energy profile of the other phonon mode transforms from a double-well potential to a single-well potential. However, the $A_3^+$ mode is found to stabilize with only a small amplitude of the $B_u$ mode. The $B_u$ phonon mode is also found to get stabilized at finite distortion resulting by $A_3^+$ mode.

The energy profiles as obtained from simultaneous excitation of $A_3^+$ with $M_3^+$ and $A_3^+$ with $M_2^-$ are shown in Fig. 8(b) and 8(c), respectively. One can see that the distortion in $M_3^+$ or $M_2^-$ modes also stablizes the unstable $A_3^+$ mode. However, the required magnitude of distortion of $M_2^-$ mode is much larger than that of $M_3^+$. In Fig. 8(d) and 8(e), we show the energy profiles as obtained from simultaneous excitation of $B_u$ mode with $M_3^+$ and $B_u$ mode with $M_2^-$ mode respectively. The distortion of $M_3^+$ leads to a single minima for the $B_u$ mode but at very high energy. On the other hand, the $M_2^-$ mode does not stabilize the zone centre $B_u$ instability as it remains a double-well even at large distortion of $M_2^-$ mode. In conclusion, the $B_u$ and the modes at M point show anharmonic coupling with the unstable mode at A point. The coupling between $A_3^+$ and $B_u$ mode leads to stabilization of $A_3^+$ mode (Fig. 8(a)) prior to $B_u$. The stabilization of $B_u$ at finite amplitude of $A_3^+$ and $M_3^+$ is not of any consequence (Fig. 8(a, d)) as it does not lead to a deeper minima in total energy. It is clear that any of the mode coupling involving only two modes as discussed above does not explain the observed distortion in $CaFeO_2$.

As mentioned above the amplitude mode analysis[33] indicates that the distortion in p-$CaFeO_2$ is induced by $B_u$, $M_3^+$ and $M_2^-$ phonon modes. The calculated structures of d-$CaFeO_2$ (P-$42_1$m) and (√2×√2×1) super cell of p-$CaFeO_2$ (P4/mmm) is given in TABLE IV. The difference in the atomic coordinates of room temperature phase (P$6_3$*cm*) and the (√2×√2×1 super cell of the p-$CaFeO_2$ (P4/mmm) is a measure of the distortion required to stabilize the d-$CaFeO_2$ (P-$42_1$m). The eigen vectors of the unstable $B_u$ and stable M point modes of p-$CaFeO_2$ (P4/mmm) for the super cell are given in Table IV. The eigen vector as obtained with appropriate weight of $B_u$ (52%) and $M_3^+$ (18%) and $M_2^-$ (30%) point modes is used to generate the observed distortion vector, which matches very well with the distortion vector. This clearly shows that the coupling between $B_{2u}$, $M_3^+$, and $M_2^-$ point modes is able to explain the observed distortion in planer structure and stabilizes d-$CaFeO_2$. Usually the cell doubling in the plane may be expected due to a soft mode at the M(½½ 0) point. However, in the present case all the modes at the M point are stable and instabilities are found at Γ and A point. We find that the stable M-point modes ($M_3^+$ and $M_2^-$) couple anharmonicity with the Γ point Bu phonon and result in the cell doubling in the a-b plane. It also turns out that the soft mode at the A-point does not have any role in inducing the distortion or the cell doubling.

We have also determined the energy barrier between the p-$CaFeO_2$ (P4/mmm) and the d-$CaFeO_2$ (P-$42_1$m) structures. We started with the calculated structure (TABLE IV) of p-$CaFeO_2$ (P4/mmm). The distortion vector as given in TABLE IV is further used to obtain (Fig 9) the profile of the energy barrier. The energy barrier between the two structures is calculated to be 0.3 eV. The calculations clearly show that a minimum in the profile is obtained at unit distortion in the p-$CaFeO_2$, thus confirms the stability of d-$CaFeO_2$.

Our ab-initio calculations for $SrFeO_2$ reveal that with increase of volume a $B_u$ mode becomes unstable. This unstable mode is similar to that found in p-$CaFeO_2$ at ambient conditions. We suggest that the distortion in $SrFeO_2$ as revealed from the phonon measurements at above 300 K is similar to that known in d-$CaFeO_2$ at ambient conditions. This is qualitatively in agreement with the nature of distortion reported[21] from an analysis of the total neutron scattering in $SrFeO_2$ at around 450 K.

### D. Phase Transition from Distorted to Planer $CaFeO_2$ at High Pressure

At high pressure, the interlayer separation between $FeO_4$ planes reduces, which may stabilize the planer structure of p-$CaFeO_2$. The comparison of the distorted and planar structure (TABLE IV) at ambient pressure shows that difference in the two structures arises due to distortion in the atomic positions of the oxygen and calcium atoms in the p-$CaFeO_2$. The calculated x and z-coordinate of the oxygen and z-coordinate of the calcium atom in the d-$CaFeO_2$ as a function of pressure is shown in Fig. 10(a), which indicates that with increase of pressure the distorted structure finally transform to the p-$CaFeO_2$ (Fig. 10(a)) at around 20 GPa. In order to check the dynamical stability of the p-$CaFeO_2$ structure, we calculated the phonon dispersion relation in the entire Brillouin zone at 30 GPa. As shown in Fig. 10(b), all the phonon modes in p-$CaFeO_2$ are indeed dynamically stable. This suggests a second order phase transition to the distorted to planar structure

### E. Spin Phonon Coupling and Magnetic Exchange Interaction Parameters in Planar $SrFeO_2$ and $CaFeO_2$

The electronic structure calculations as reported in the literature[34] show a possibility of electromagnetic coupling in planar $BaFeO_2$ as well as distorted $CaFeO_2$. In addition, a significant change was shown in electronic contribution to the total density of states of different atoms with different magnetic configurations (A, C, G antiferromagnetic and ferromagnetic F). In A-type antiferromagnetic configuration Fe atoms interact ferromagnetically within a-b plane and antiferromagnetically to adjacent a-b plane. In C-type configuration the intraplaner interaction between Fe atoms is antiferromagnetic in nature but interplaner interaction is ferromagnetic. However in G-type configuration all Fe atoms interact antiferromagnetically with nearest Fe atoms. This motivated us to perform the phonon calculations of p-$CaFeO_2$ and $SrFeO_2$ in various antiferromagnetic magnetic configurations, namely, A, C, G as well as in the ferromagnetic (F) configurations. As shown in Fig. 11, we could not observe significant difference in the calculated phonon dispersion relations for p-$CaFeO_2$ in various magnetic configurations. The zone-centre and zone-boundary instabilities remained present in all the calculations while their magnitude of instabilities changed slightly. This shows that the p-$CaFeO_2$ structure is not dynamically stable in the above said magnetic structures.

As mentioned above, $SrFeO_2$ is found dynamically stable with the G-type AFM ordering. We find that zone centre $B_u$ mode for $SrFeO_2$ becomes unstable in FM and A-type spin configurations, while in C-type and G-type it is dynamically stable. In FM and A type configurations, the parallely aligned Fe moments within the layer result in a slightly larger Fe-O bond length (2.03 Å) in comparison to the Fe-O bond (2.02 Å) in C and G type configurations with anti-parallelly aligned Fe moments (see Table IV).

The phonon spectra are also sensitive to the interlayer distances in the quasi two-dimensional systems.The calculations performed including the C type and G-Type structures on $SrFeO_2$ seem to result in a slightly larger interlayer separation (TABLE V), in comparison to the A- and F-type configurations. Figure 11 shows that a small change in interlayer distance caused by different magnetic interaction gives significant influences on phonon spectra. These calculations suggest that nature of magnetic configurations have significant impact on the structural stability as well as anharmonicity of phonons. As mentioned above, the phonon spectra in $SrFeO_2$ change significantly with the change in magnetic configurations, implying a strong spin phonon coupling. However, phonons in p-$CaFeO_2$ show a weak dependence with change in magnetic configurations, indicating a weak spin-phonon coupling of phonons.

Analysis of the high-resolution neutron diffraction measurements on $SrFeO_2$ shows[21] that the exchange parameters are reduced significantly above 300 K along with the local distortion in the planer geometry. We calculated the exchange interaction parameters $J_1$, $J_2$, $J_3$ and $J_4$ for $SrFeO_2$ and $CaFeO_2$ (see Fig. 12(a)). The calculated exchange parameters at different amplitudes of phonon distortion of $B_u$ mode (Fig. 12(b))show that with the increase in the distortion amplitude of $B_u$ phonon mode, the exchange parameters indeed reduce significantly, in consistent with the experiment[35]. At high temperature, the amplitude and population of low energy modes should be significantly large and hence the distortion involving out-of-plane oxygen motion (Fig. 7) would be enhanced.

**F. Comparison of the Phonon Spectra of $SrFeO_2$ and $SrFeO_3$**

The $AMO_3$ compounds exhibit a wide variation in their electrical, magnetic and other physicochemical properties with respect to the composition, temperature, etc. They also exhibit significant changes in their properties when the M or A cation is changed from one element to other. Goodenough[36] has given a qualitative chemical description of these perovskite oxides in terms of the M-O overlap parameter and has classified a large number of these oxides according to their M-cation spin and the M-O overlap strength. The calculations based on DFT[35] have been performed to understand the different band parameters such as the Fermi energy, density of states at the Fermi energy, conduction-band width, etc. However calculations are not available in the literature which highlight the effect of M-O-M linkage in the three and two dimensional systems; which is supposed to affect the phonon spectra and hence the thermodynamical properties.

The Fe atoms in $SrFeO_3$ are six-fold coordinated and forms $FeO_6$ regular octahedra. At ambient conditions $SrFeO_3$ has the G type spin structure and crystallizes in cubic phase (*Pm*-3*m*). The neutron measurements[37] show that the spin structure is not collinear and has a helical form. Computationally phonon calculations with a helical magnetic structure are very cumbersome. So in order to see the stability of $SrFeO_3$,we have calculated the total energy of $SrFeO_3$ including the A, C, G type collinear structures as well in the FM structure. We find that ferromagnetic structure is the most stable magnetic configuration in agreement with the literature[38]. We have calculated the phonon spectra (Fig. 13) of $SrFeO_3$ including the ferromagnetic structure and compared it with that of $SrFeO_2$, where Fe has square planar co-ordination. We find large difference in the phonon spectra of both compounds.

The cubic unit cell of $SrFeO_3$ is slightly larger volume in comparison to $SrFeO_2$. The Fe-O bond of 1.95 Å in $SrFeO_3$ is shorter as compared to 2.0 Å in $SrFeO_2$. However, Sr-O/Sr-Fe bonds in $SrFeO_3$ (2.75/3.37 Å) are slightly larger in comparison to that in $SrFeO_2$ (2.64/3.31 Å). The contributions due to Sr atoms in both the compounds extend (Fig. 13) up to about 30 meV. The partial density of states calculations show that Sr atoms in both the compounds have peaks at 15 meV. However Sr vibrations in $SrFeO_3$(cubic) are more localized to lower energies, which may be attributed to the its high symmetry in comparison to $SrFeO_2$ (tetragonal).The contribution from oxygen's in both the compound are in the entire spectral range. The dynamics of oxygen atoms in $SrFeO_2$ extends to high energies in comparison to that in $SrFeO_3$.This is attributed to strong planer Fe-O boding in $SrFeO_2$. The vibrations due to Fe atoms in $SrFeO_3$ seem to extend to 50 meV in comparison to 40 meV in $SrFeO_2$.The Fe-O stretching modes in $SrFeO_3$ are up to 70 meV in comparison to 80 meV in $SrFeO_2$.The fact that the modes span the lower energy range in $SrFeO_3$ indicates the difference in the nature of bonding in the two compounds.

## IV. CONCLUSIONS

We have reported detailed measurements of the temperature dependence of the phonon density-of-states of $SrFeO_2$ in the antiferromagnetic phase ($P4/mmm$). Anharmonic broadening above 300 K may be related to the reported distortion of the planer structure.The phonon spectra of $SrFeO_2$ have been analysed based on detailed *ab-initio* lattice dynamical calculations in the magnetic state. Our calculations show that Fe magnetism should be considered for obtaining the dynamically stable *P4/mmm* structure of $SrFeO_2$. However the calculations carried out in the same space group for $CaFeO_2$ result in dynamically unstable structure. The lattice contraction in $CaFeO_2$ in compare to $SrFeO_2$ does not explain the presence of phonon instabilities, which may be due to difference in nature of bonding in $SrFeO_2$ and $CaFeO_2$. The anharmonic coupling of the unstable $B_u$ mode and stable zone boundary modes at M point may lead to the distortion in the planer structure (P4/mmm) and may be responsible for stabilization of d-$CaFeO_2$ ($P$-$42_1m$). These observations are consistent with the available experimental structural data. The undistorted planer $CaFeO_2$ becomes dynamically stable at high pressures. The spin exchange interaction parameters in the P4/mmm space group are found to decrease with increase in the distortion of the structure as described by the amplitude of $B_u$ phonon mode.

TABLE I. Comparison of the calculated structural parameters of $SrFeO_2$ and $CaFeO_2$ with the experimental data. The experimental data [6, 10] of lattice parameters is at 293 K, while the calculated values are given at 0 K. For isotropic temperature factors experimental data [14] and calculations are given at 293 K.

| | $SrFeO_2$, *P*4/*mmm* Expt. [3] | $SrFeO_2$ *P*4/*mmm* Calc. | $CaFeO_2$, *P*-$42_1m$ Expt [6] | $CaFeO_2$ *P*-$42_1m$ Calc. |
|---|---|---|---|---|
| *a*(Å) | 3.991 | 4.042 | 5.507 | 5.550 |
| *c*(Å) | 3.474 | 3.497 | 3.355 | 3.443 |
| $B_{iso}$(Sr/Ca) Å$^2$ | 0.470 | 0.440 | 0.485 | 0.580 |
| $B_{iso}$(Fe) Å$^2$ | 0.470 | 0.380 | 0.590 | 0.490 |
| $B_{iso}$(O) Å$^2$ | 0.790 | 0.610 | 0.909 | 0.660 |
| Volume/ atom | 13.83 | 14.28 | 12.72 | 13.26 |

TABLE II(a). The calculated zone centre optic phonon modes for $SrFeO_2$ (*P*4/*mmm*) and d-$CaFeO_2$ (*P*-$42_1m$) in meV units. (1 meV=8.0585 cm$^{-1}$).

| | $SrFeO_2$ (*P*4/*mmm*) | $CaFeO_2$ (*P*4/*mmm*) | $CaFeO_2$ (*P*-$42_1m$) | |
|---|---|---|---|---|
| $A_u$ | 23.7<br>44.7 | 24.9<br>44.4 | $A_1$ | 15.2<br>26.9<br>35.7 |
| $B_u$ | 9.4 | 13.7*i* | $A_2$ | 65.4 |
| $E_u$ | 21.2<br>35.0<br>62.6 | 23.1<br>33.9<br>71.4 | $B_1$ | 20.2<br>44.6 |
| | | | $B_2$ | 25.2<br>48.9<br>60.5 |
| | | | E | 16.0<br>27.8<br>29.7<br>38.9<br>41.6 |

TABLE II (b). The calculated zone boundary modes for $CaFeO_2$ (*P*4/*mmm*) in meV units. (1 meV=8.0585 $cm^{-1}$). $M_5^+$, $M_5^-$, $A_5^+$ and $A_4^-$ are doubly degenerate modes.

| $CaFeO_2$ (*P*4/*mmm*) | | | |
|---|---|---|---|
| $M_3^+$ | 8.2 | $A_3^+$ | 11.5*i* |
| $M_5^+$ | 12.6 | $A_5^+$ | 17.9 |
| $M_2^-$ | 13.7 | $A_3^-$ | 19.9 |
| $M_5^-$ | 19.1 | $A_4^+$ | 29.5 |
| $M_3^-$ | 21.9 | $A_5^+$ | 30.4 |
| $M_5^-$ | 43.0 | $A_5^-$ | 36.8 |
| $M_2^+$ | 48.6 | $A_2^+$ | 46.4 |
| $M_4^+$ | 56.7 | $A_4^+$ | 50.2 |
| $M_1^+$ | 73.1 | $A_1^+$ | 68.6 |

TABLE III. Calculated Born effective charges (Z) as well as dielectric constants (ε) in various phases of $SrFeO_2$ and $CaFeO_2$.

| | $Z_{xx}$ | $Z_{xy}$ | $Z_{yx}$ | $Z_{yy}$ | $Z_{zz}$ |
|---|---|---|---|---|---|
| $SrFeO_2$(*P*4/*mmm*) $\varepsilon_{xx}$ = 5.02, $\varepsilon_{zz}$ = 5.82 | | | | | |
| Sr | 2.29 | 0 | 0 | 2.29 | 2.97 |
| Fe | 2.97 | 0 | 0 | 2.97 | 0.35 |
| O | -3.16 | 0 | 0 | -2.10 | -1.66 |
| p-$CaFeO_2$(*P*4/*mmm*) $\varepsilon_{xx}$ = 5.80, $\varepsilon_{zz}$ = 6.35 | | | | | |
| Ca | 2.35 | 0 | 0 | 2.35 | 2.78 |
| Fe | 2.82 | 0 | 0 | 2.82 | 0.63 |
| O | -3.03 | 0 | 0 | -2.14 | -1.71 |
| d-$CaFeO_2$ (*P*-4$2_1$*m*) $\varepsilon_{xx}$ = 4.97, $\varepsilon_{zz}$ = 4.94 | | | | | |
| Ca | 2.26 | -0.07 | -0.07 | 2.26 | 2.65 |
| Fe | 2.79 | -0.13 | 0.13 | 2.79 | 0.74 |
| O | -2.53 | 0.41 | 0.41 | -2.53 | -1.69 |

TABLE IV. The calculated structures of d-$CaFeO_2$ (P-$42_1$m) and (√2×√2×1) super cell of p-$CaFeO_2$ (P4/mmm). The super cell (√2×√2×1) of p-$CaFeO_2$ is equivalent to the d-$CaFeO_2$. The distortion vector is obtained from the difference in atomic co-ordinates between d-$CaFeO_2$ and p-$CaFeO_2$ structures phases. The eigen vector of the unstable $B_u$ and stable $M_3^+$ and $M_2^-$ modes in the p-$CaFeO_2$ phase for the super cell is also given. A linear combination of Bu , $M_3^+$ and $M_2^-$ modes with appropriate weight factor is used to generate the observed distortion vector as given in the last column of the table.

| | | **P-$42_1$m $CaFeO_2$** | **P4/mmm $CaFeO_2$** | **Distortion vector** | **$B_u$** | **$M_3^+$** | **$M_2^-$** | **0.25Bu+0.085 $M_3^+$+0.141$M_2^-$** |
|---|---|---|---|---|---|---|---|---|
| | *a (Å)* | 5.550 | 5.622 | | | | | |
| | *b (Å)* | 5.550 | 5.622 | | | | | |
| | *c (Å)* | 3.443 | 3.234 | | | | | |
| | | | | | | | | |
| O1 | *x* | 0.720 | 0.750 | 0.030 | 0.000 | 0.354 | 0.000 | 0.030 |
| | *y* | 0.780 | 0.750 | -0.030 | 0.000 | -0.354 | 0.000 | -0.030 |
| | *z* | -0.120 | 0.000 | 0.120 | 0. 500 | 0.000 | 0.000 | 0.120 |
| | | | | | | | | |
| O2 | *x* | 0.280 | 0.250 | -0.030 | 0.000 | -0.354 | 0.000 | -0.030 |
| | *y* | 0.220 | 0.250 | 0.030 | 0.000 | 0.354 | 0.000 | 0.030 |
| | *z* | -0.120 | 0.000 | 0.120 | 0.500 | 0.000 | 0.000 | 0.120 |
| | | | | | | | | |
| O3 | *x* | 0.220 | 0.250 | 0.030 | 0.000 | 0.354 | 0.000 | 0.030 |
| | *y* | 0.720 | 0.750 | 0.030 | 0.000 | 0.354 | 0.000 | 0.030 |
| | *z* | 0.120 | 0.000 | -0.120 | -0.500 | 0.000 | 0.000 | -0.120 |
| | | | | | | | | |
| O4 | *x* | 0.780 | 0.750 | -0.030 | 0.000 | -0.354 | 0.000 | -0.030 |
| | *y* | 0.280 | 0.250 | -0.030 | 0.000 | -0.354 | 0.000 | -0.030 |
| | *z* | 0.120 | 0.000 | -0.120 | -0.500 | 0.000 | 0.000 | -0.120 |
| | | | | | | | | |
| Ca1 | *x* | 0.500 | 0.500 | 0.000 | 0.000 | 0.000 | 0.000 | 0.000 |
| | *y* | 0.000 | 0.000 | 0.000 | 0.000 | 0.000 | 0.000 | 0.000 |
| | *z* | 0.400 | 0.500 | 0.100 | 0.000 | 0.000 | 0.707 | 0.100 |
| | | | | | | | | |
| Ca2 | *x* | 0.000 | 0.000 | 0.000 | 0.000 | 0.000 | 0.000 | 0.000 |
| | *y* | 0.500 | 0.500 | 0.000 | 0.000 | 0.000 | 0.000 | 0.000 |
| | *z* | 0.600 | 0.500 | -0.100 | 0.000 | 0.000 | -0.707 | -0.100 |
| | | | | | | | | |
| Fe1 | *x* | 0.000 | 0.000 | 0.000 | 0.000 | 0.000 | 0.000 | 0.000 |
| | *y* | 0.000 | 0.000 | 0.000 | 0.000 | 0.000 | 0.000 | 0.000 |
| | *z* | 0.000 | 0.000 | 0.000 | 0.000 | 0.000 | 0.000 | 0.000 |
| | | | | | | | | |
| Fe2 | *x* | 0.500 | 0.500 | 0.000 | 0.000 | 0.000 | 0.000 | 0.000 |
| | *y* | 0.500 | 0.500 | 0.000 | 0.000 | 0.000 | 0.000 | 0.000 |
| | *z* | 0.000 | 0.000 | 0.000 | 0.000 | 0.000 | 0.000 | 0.000 |

TABLE V. Calculated lattice parameters and bond length of $CaFeO_2$ and $SrFeO_2$ in various magnetic configurations (A, C, G type antiferromagnetic and Ferromagnetic F).

| | $CaFeO_2$ | | | | $SrFeO_2$ | | | |
|---|---|---|---|---|---|---|---|---|
| | G | C | A | F | G | C | A | F |
| *a*(Å) | 3.975 | 3.976 | 3.994 | 3.993 | 4.042 | 4.042 | 4.061 | 4.061 |
| *b*(Å) | 3.975 | 3.976 | 3.994 | 3.993 | 4.042 | 4.042 | 4.061 | 4.061 |
| *c*(Å) | 3.234 | 3.238 | 3.229 | 3.232 | 3.497 | 3.499 | 3.487 | 3.488 |
| Fe-O(Å) | 1.987 | 1.988 | 1.997 | 1.996 | 2.021 | 2.021 | 2.031 | 2.031 |
| Interlayer Separation(Å) | 3.234 | 3.238 | 3.229 | 3.232 | 3.497 | 3.499 | 3.487 | 3.488 |

FIG. 1. (Color online) Structures of planer $CaFeO_2$ (*P*4/*mmm*) and distorted $CaFeO_2$(*P*-42$_1$*m*). The *ab* plane in these structure are depicted by violate sheet. Supercell's compatible to the magnetic unit cell are shown, i.e. a √2× √2 × 2 supercell of the P4/mmm structure and 1×1×2 supercell of P-42$_1$m structure.The oxygen atoms in the distorted structure are shifted along z axis by ±δ. Key: Ca, blue spheres; Fe, golden spheres; O, red spheres.

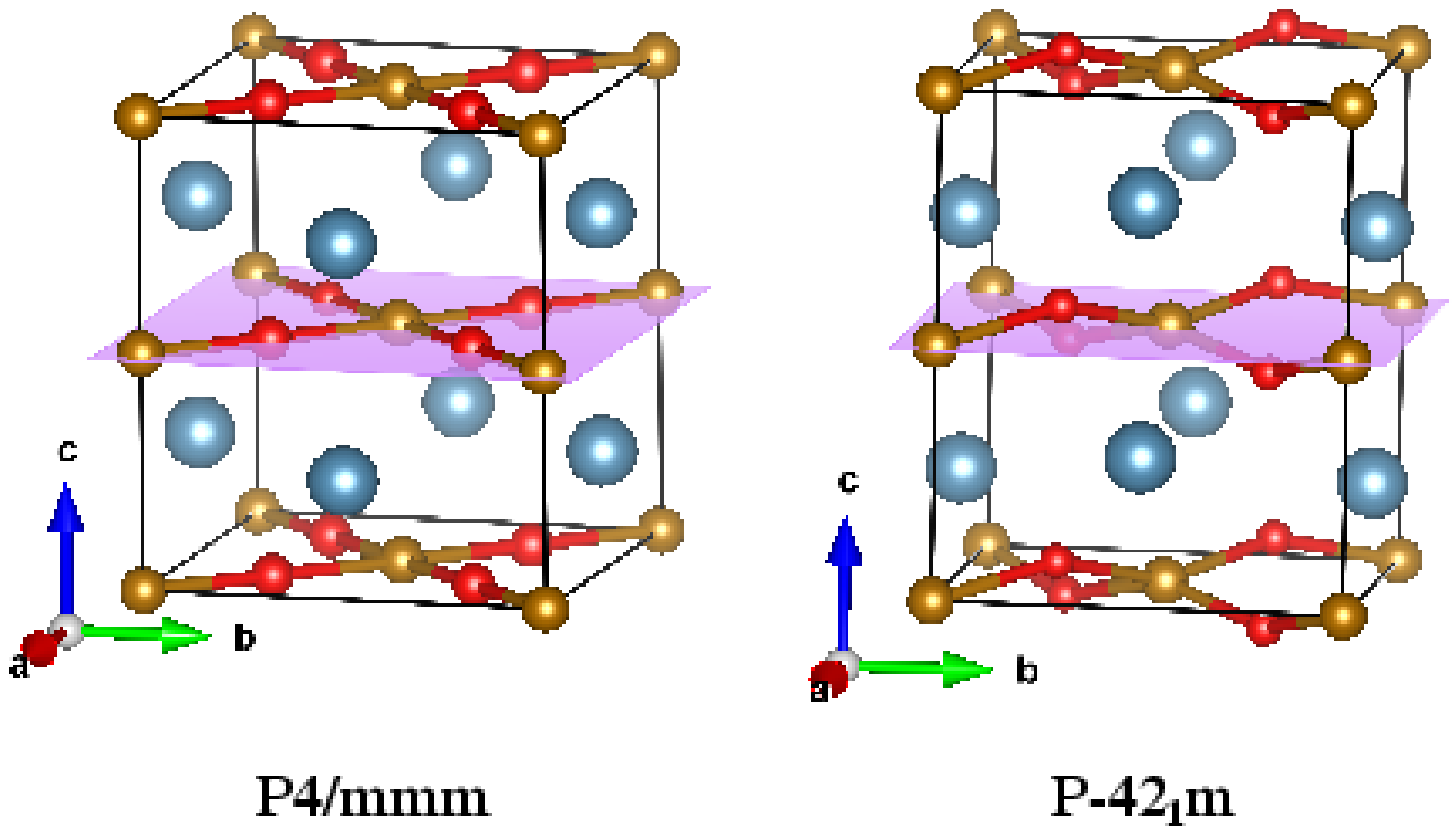

FIG. 2. (Color online) Experimental phonon spectra of $SrFeO_2$ (*P*4/*mmm*) at various temperatures in the antiferromagnetic phase.

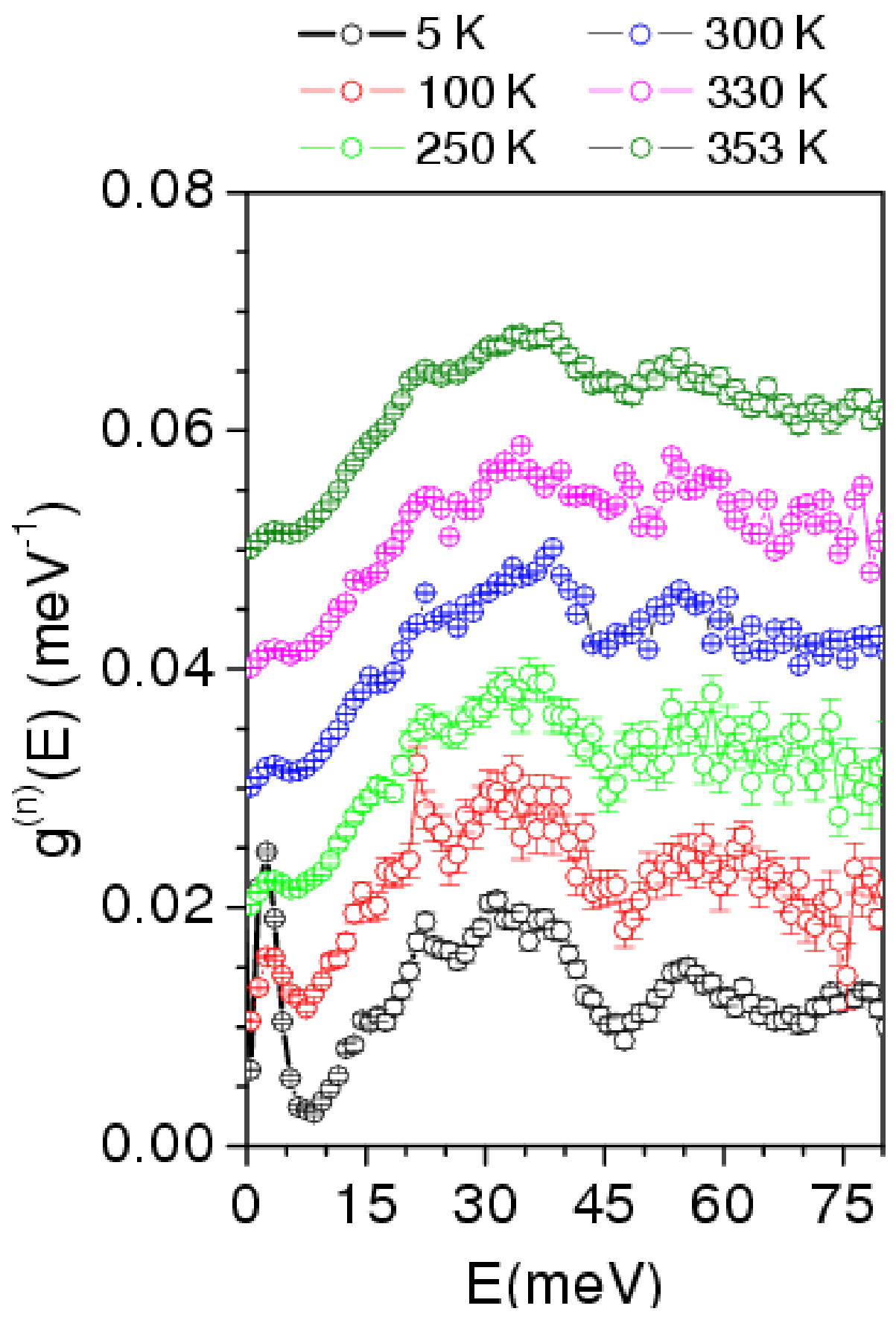

FIG. 3. Experimental and calculated phonon spectra of $SrFeO_2$ (*P*4/*mmm*).The partial atomic contributions to total neutron weighted phonon density of states are shown with dotted lines. The calculated spectra have been convoluted with a Gaussian of FWHM of 7meV of the energy transfer in order to describe the effect of energy resolution in the experiment. In order to compare with the experimental data the calculated spectrum is scaled by 6%.

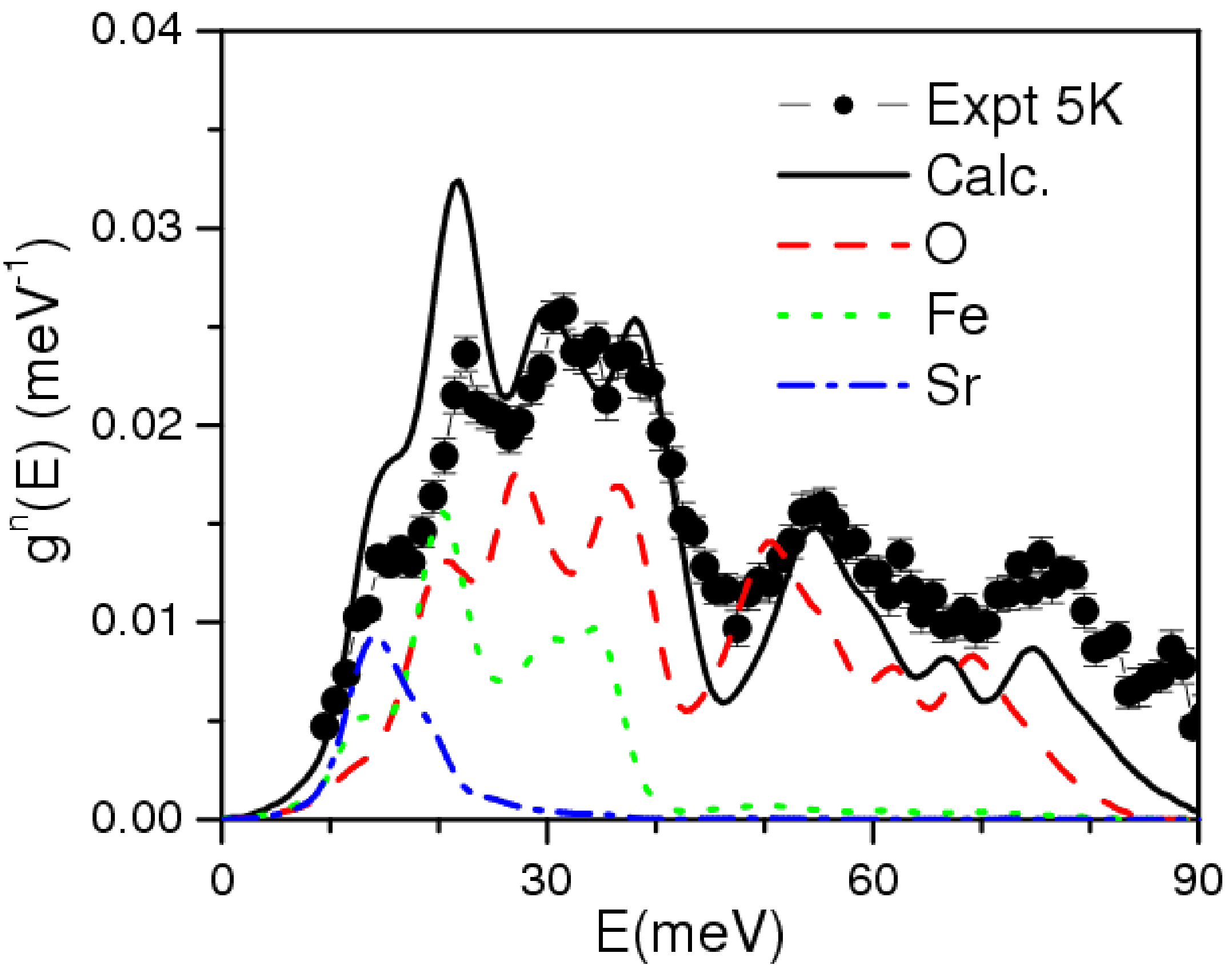

FIG. 4. (Color online) Calculated partial densities of states of various atoms in $SrFeO_2$(*P*4/*mmm*),$CaFeO_2$ (*P*4/*mmm*) and $CaFeO_2$(*P*-42$_1$*m*).

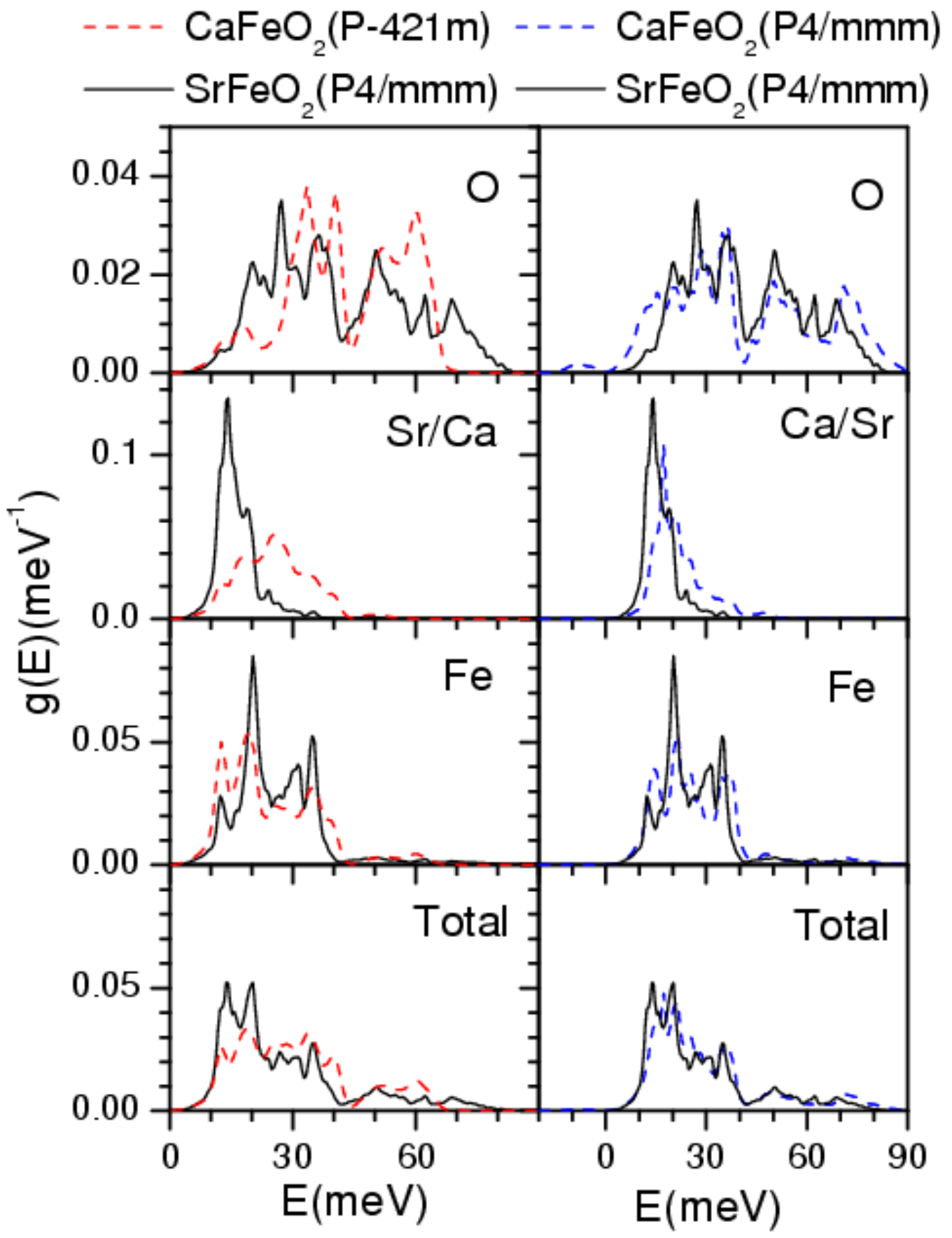

FIG. 5. (Color online) Calculated mean square amplitude ($u^2$) of various atoms in $SrFeO_2$ (*P*4/*mmm*) and d-$CaFeO_2$ (*P*-42$_1$*m*) as a function of temperature. The *x* and *z* components of $u^2$ for an atom correspond to in-plane and out-of-plane components, respectively.

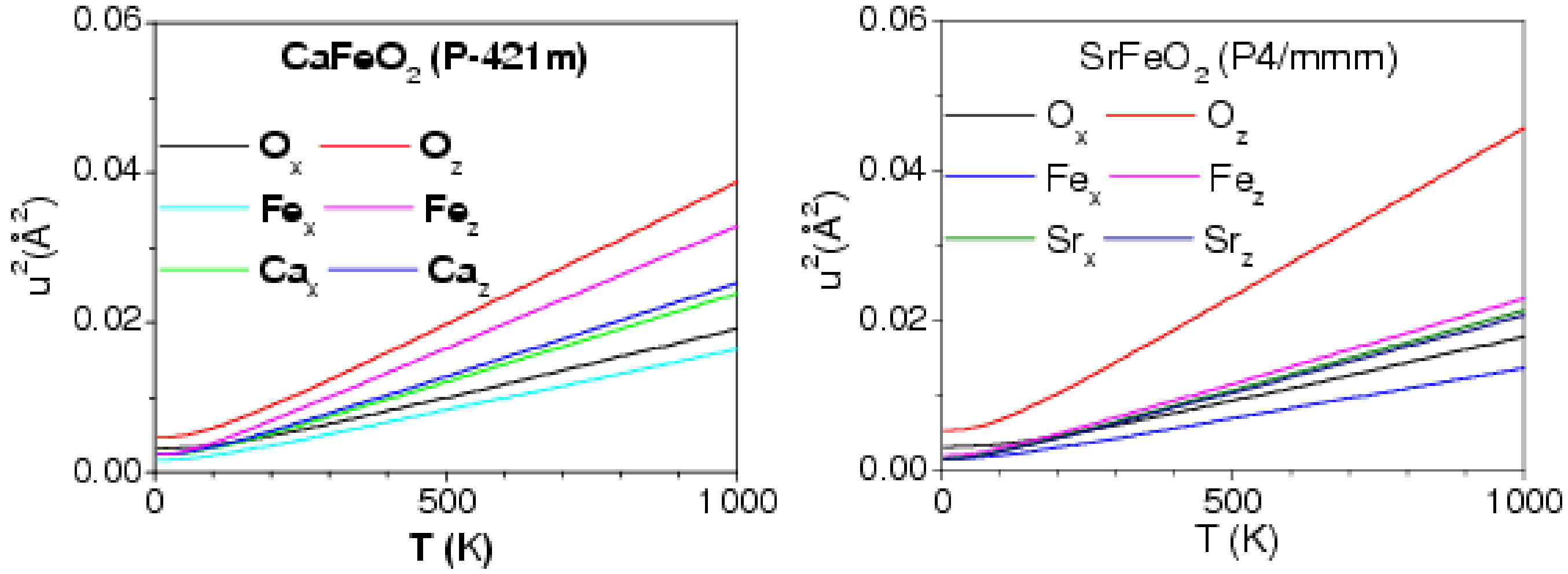

FIG. 6.(Color online) Calculated dispersion relation of $SrFeO_2$ and $CaFeO_2$in P4/mmm space group.. The solid and dashed lines correspond to calculations at ambient pressure and 5 kbar, respectively. The Bradley-Cracknell notation is used for the high symmetry points along which the dispersion relations are obtained: Γ=(0, 0, 0), Z = (1/2, 0, 0) , M = (1/2, 1/2 , 0). A = (1/2, 1/2, 1/2), R = (0, 1/2, 1/2), X = (1/2, 0, 0).

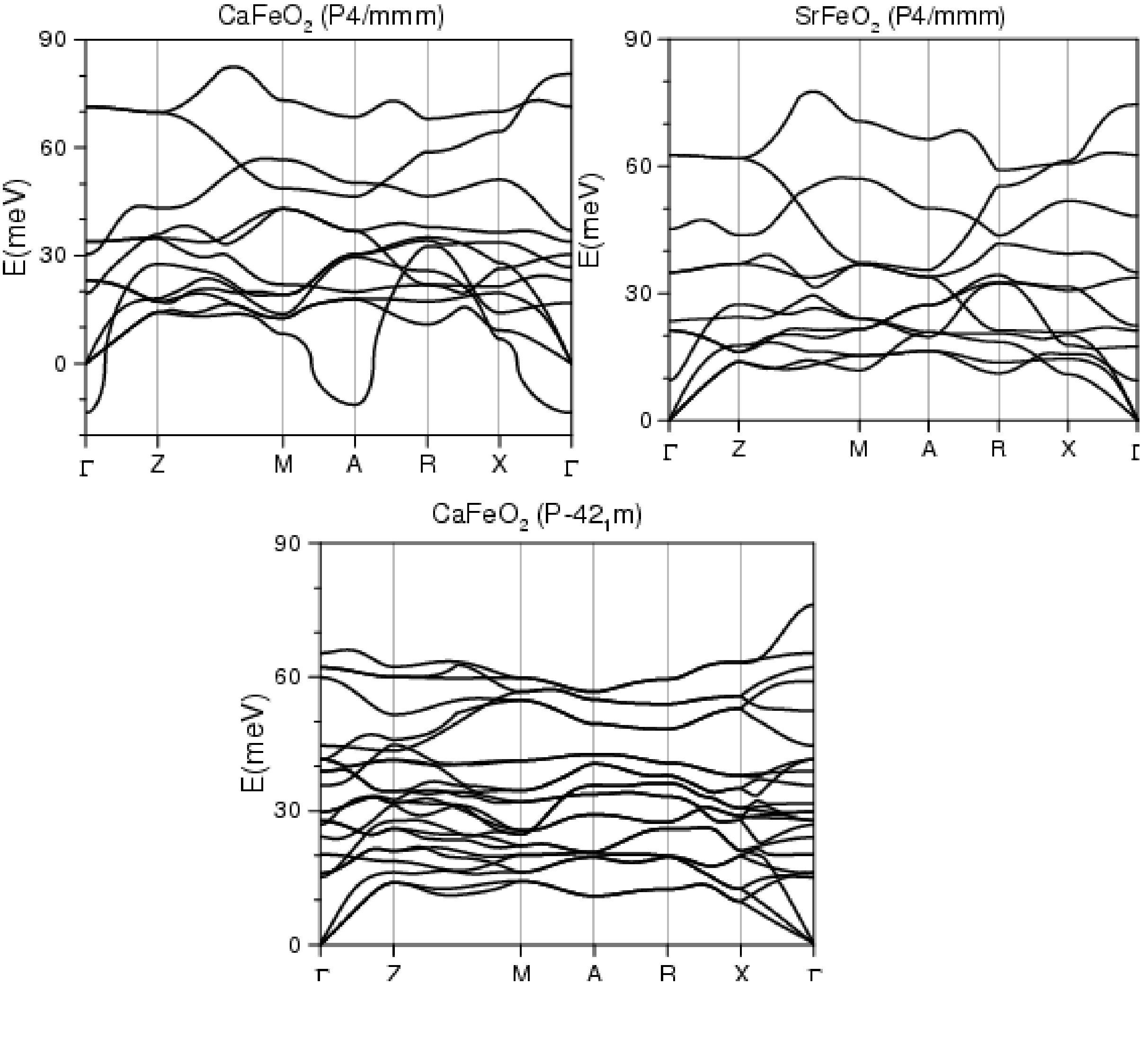

FIG. 7. (Color online) Polarization vectors of selected zone center modes of $CaFeO_2$in *P*4/*mmm*For each mode, the assignment and frequency are indicated in meV units. The *'i'* after the phonon energy indicates that mode is unstable. The length of the arrows is related to the displacement of the atoms. The absence of an arrow on an atom indicates that the atom is at rest. The number after the mode assignment gives the phonon frequency. Key: Ca, blue spheres; Fe, golden spheres; O, red spheres (1 meV=8.0585 $cm^{-1}$).

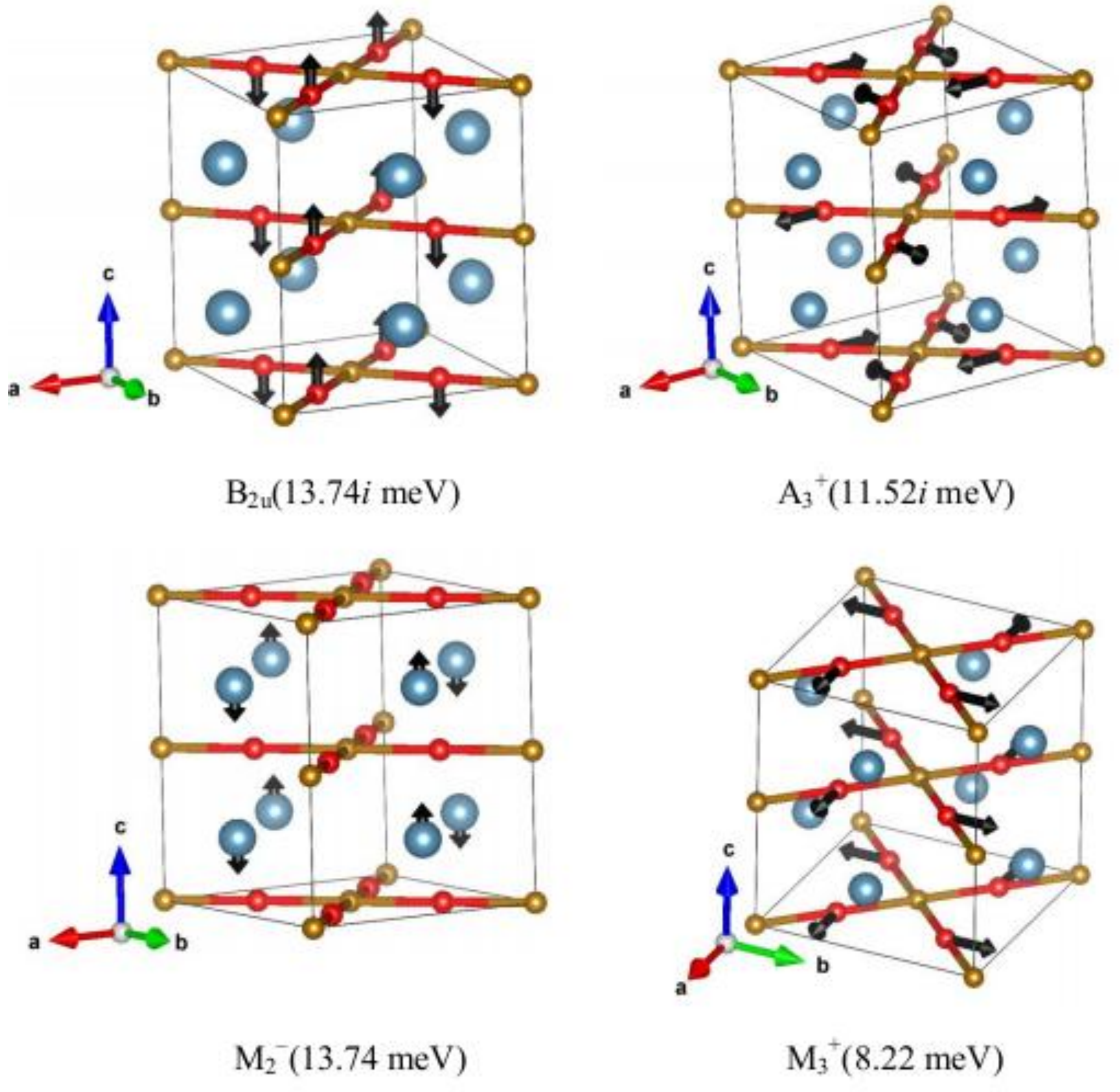

FIG. 8. (Color online) The energy landscape of p-$CaFeO_2$ obtained by exciting the pair of phonons with different amplitude. (a) $A^{3+}$and $B_u$ modes (b) $M^{3+}$ and $A^{3+}$ modes (c) $M^{2-}$ and$A^{3+}$ modes (d) $M^{3+}$and $B_u$ modes and (e) $M^{2-}$and $B_u$ modes. The energies E are per magnetic unit cell.

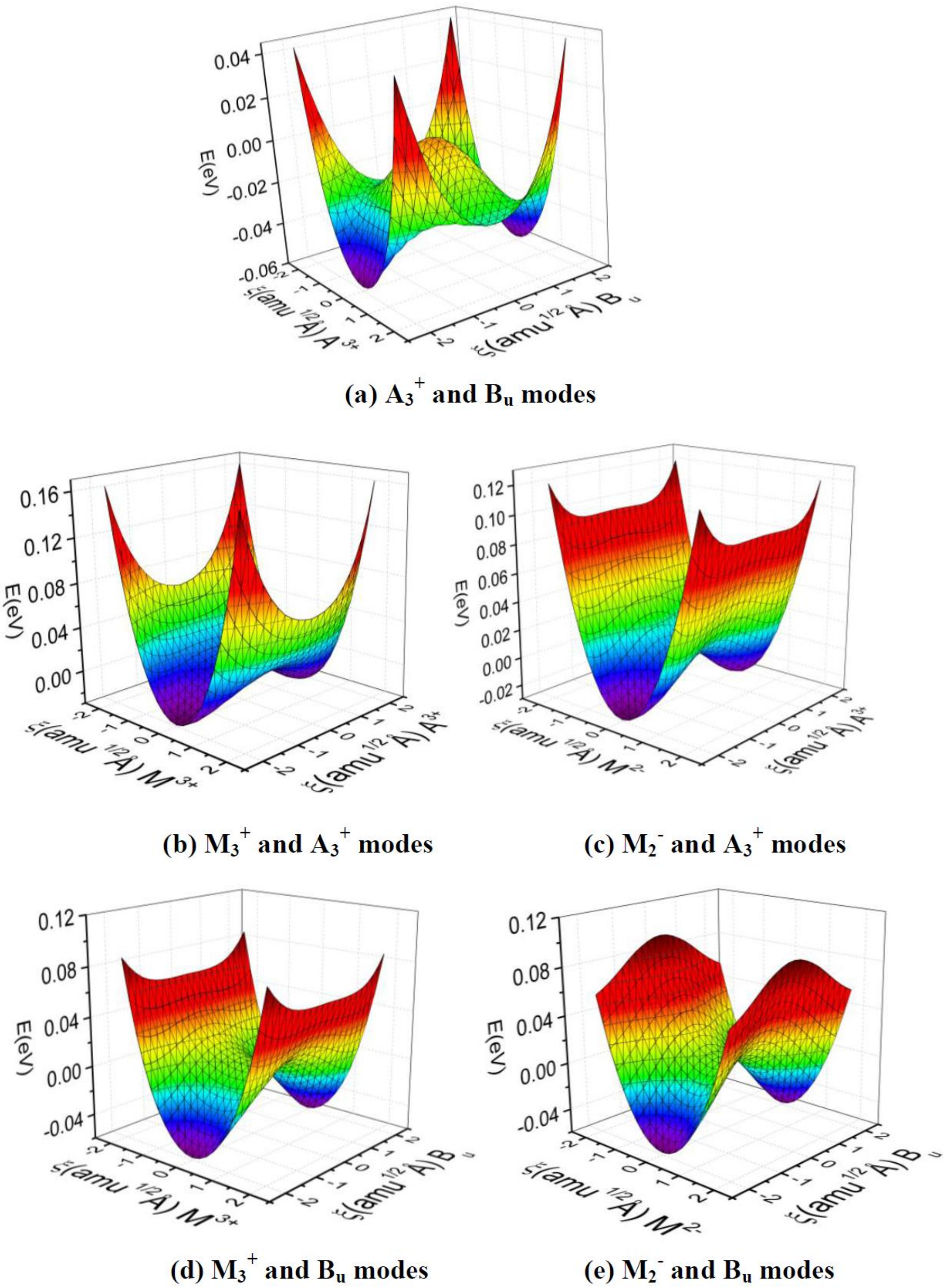

FIG. 9. Energy barrier from the p-$CaFeO_2$ (P4/mmm) to the d-$CaFeO_2$ (P-$42_1$m). ζ corresponds to the distortion vector as obtained from the difference in atomic co-ordinates of the d-$CaFeO_2$ and p-$CaFeO_2$ structures phases as given in TABLE IV. The energies E is per magnetic unit cell.

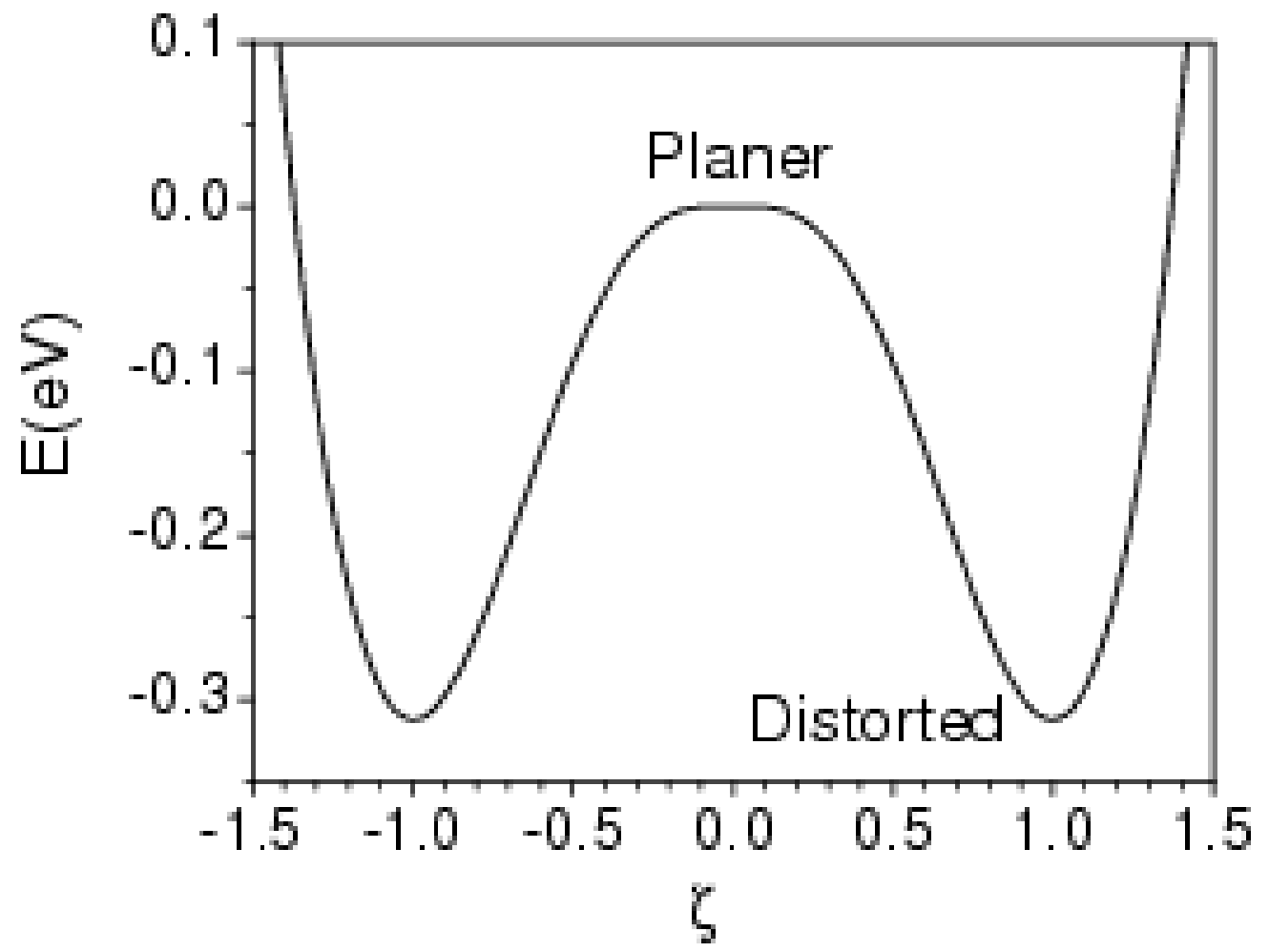

FIG. 10. (Color online) (a) The calculated x and z-coordinate of the oxygen and z-coordinate of the calcium atom in the d-$CaFeO_2$ as a function of pressure. As given in Table IV the oxygen and calcium atoms occupy the Wyckoff sites 4e(x+1/2, -x, -z) and (1/2 0 –z) respectively. (b) The calculated phonon dispersion of planer $CaFeO_2$ at ambient and 30GPa.

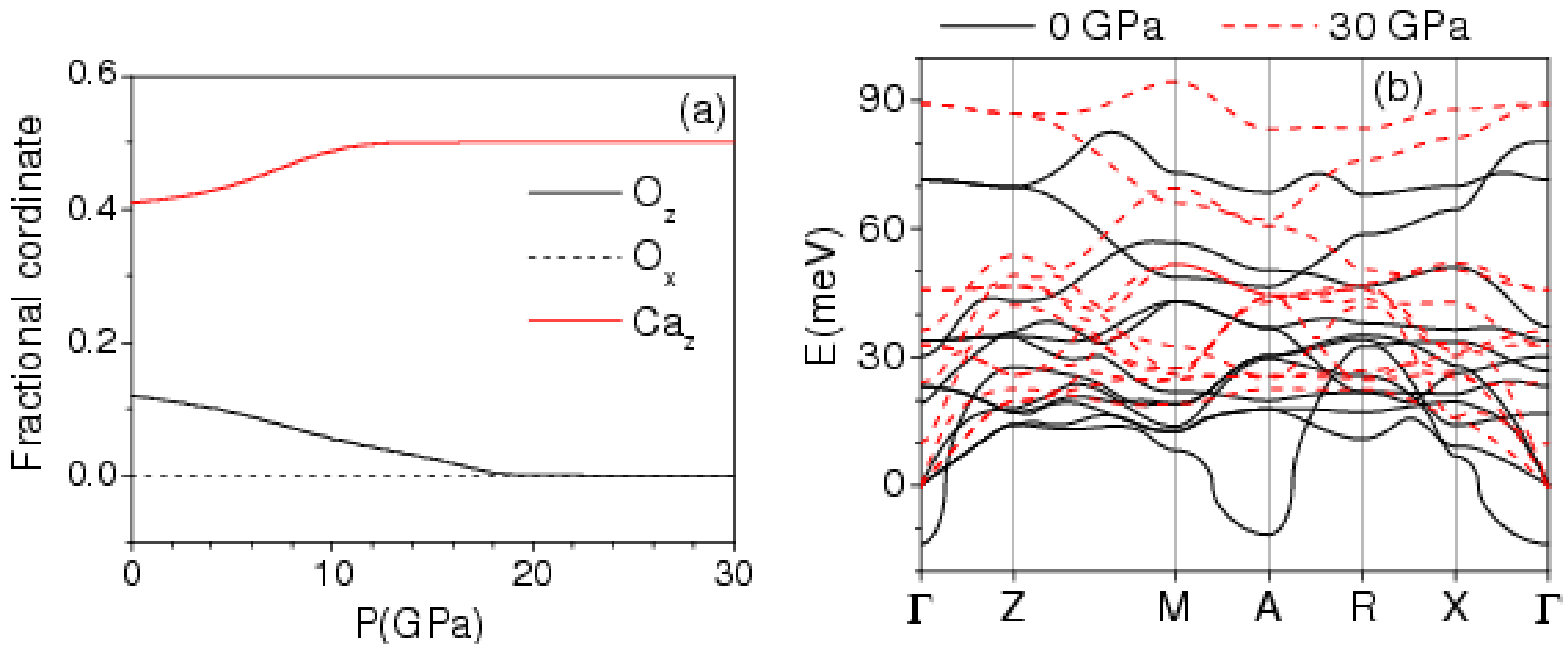

FIG 11. Calculated dispersion relations of $SrFeO_2$(*P*4/*mmm*) and $CaFeO_2$(*P*4/*mmm*)including the A, C, G antiferromagnetic and FM configurations.

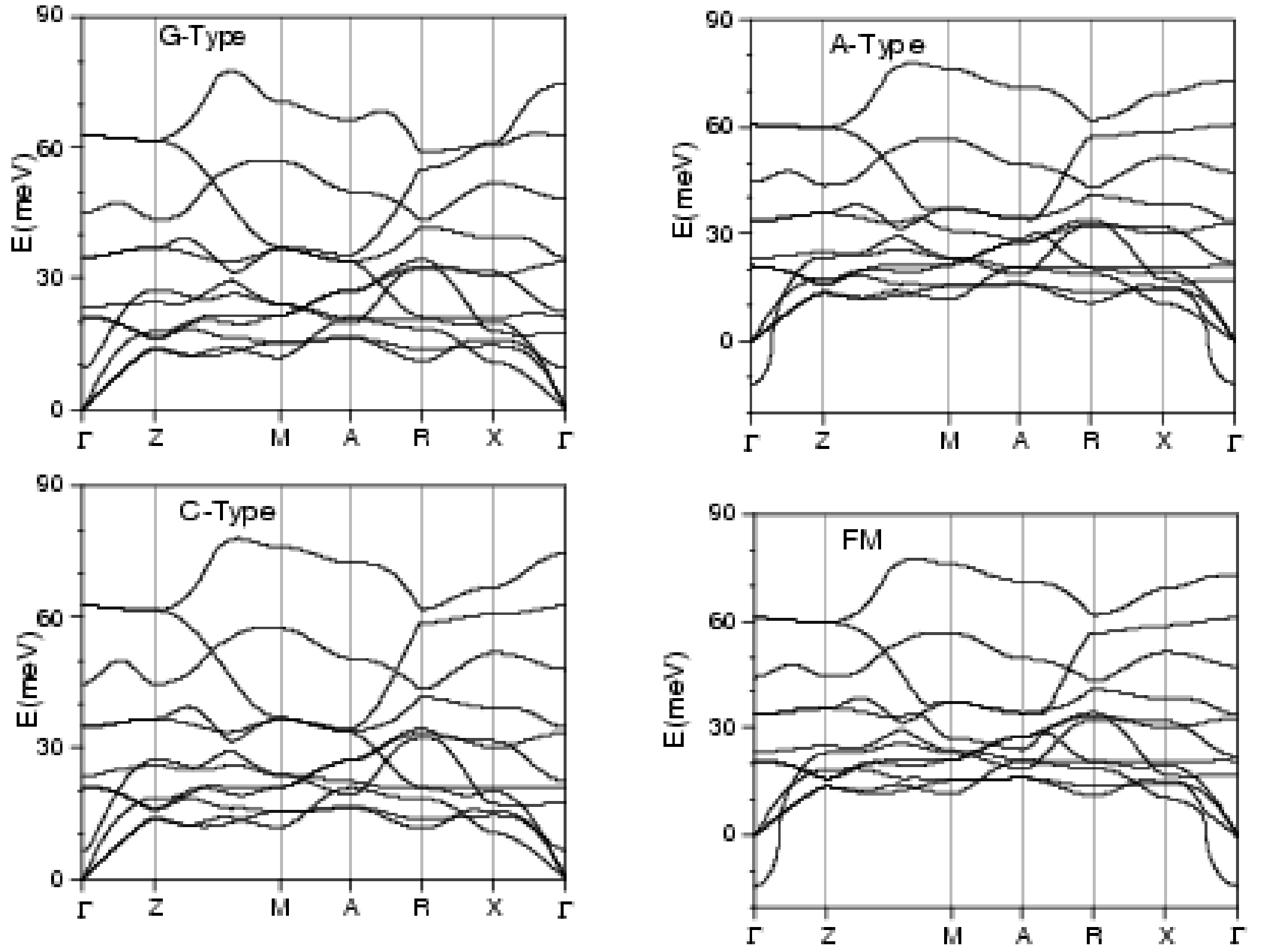

**$CaFeO_2$**

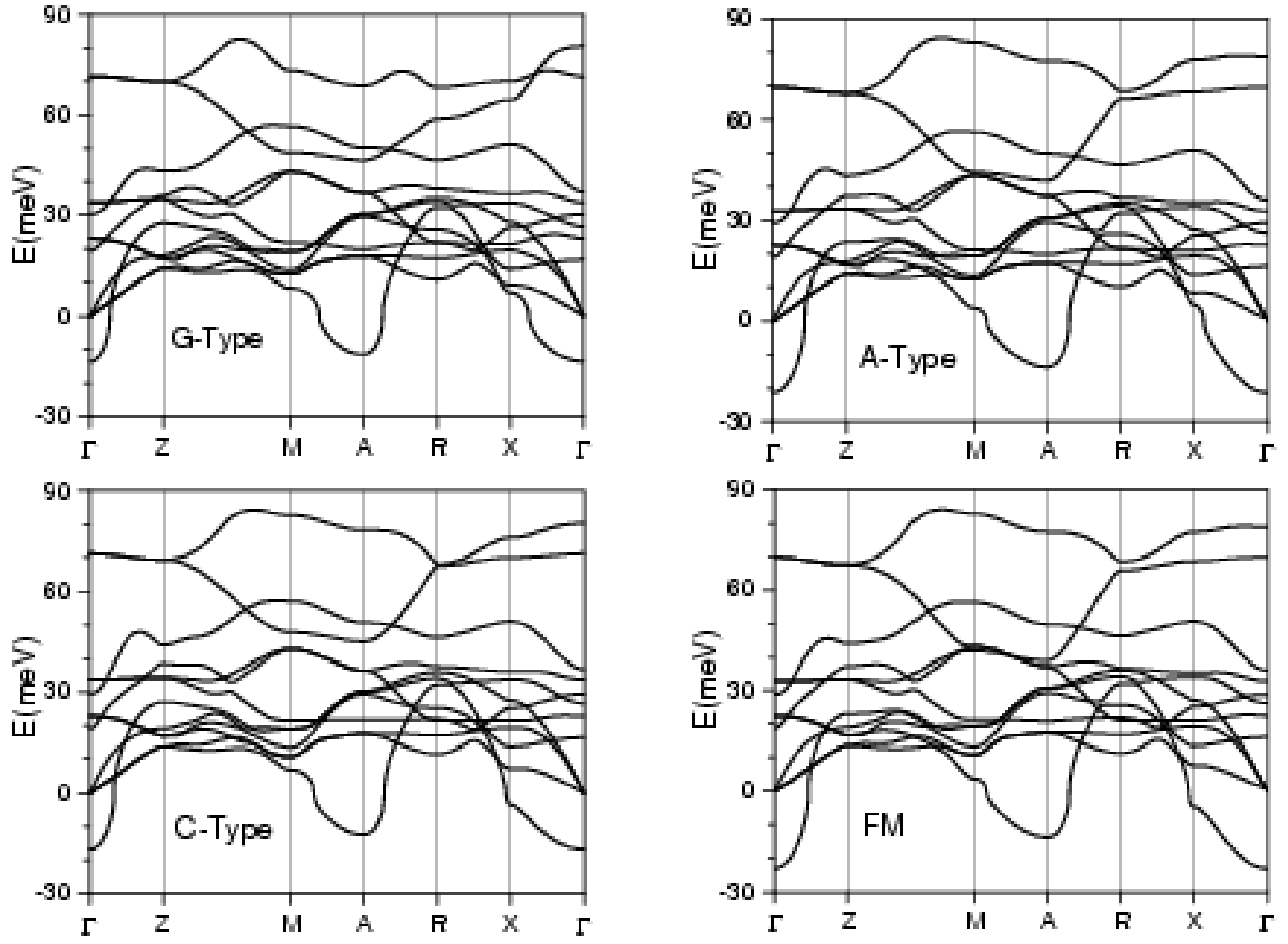

FIG. 12 (a) (Color online) The various J exchange interaction parameters in $SrFeO_2$ (*P*4/*mmm*) and p-$CaFeO_2$(*P*4/*mmm*).

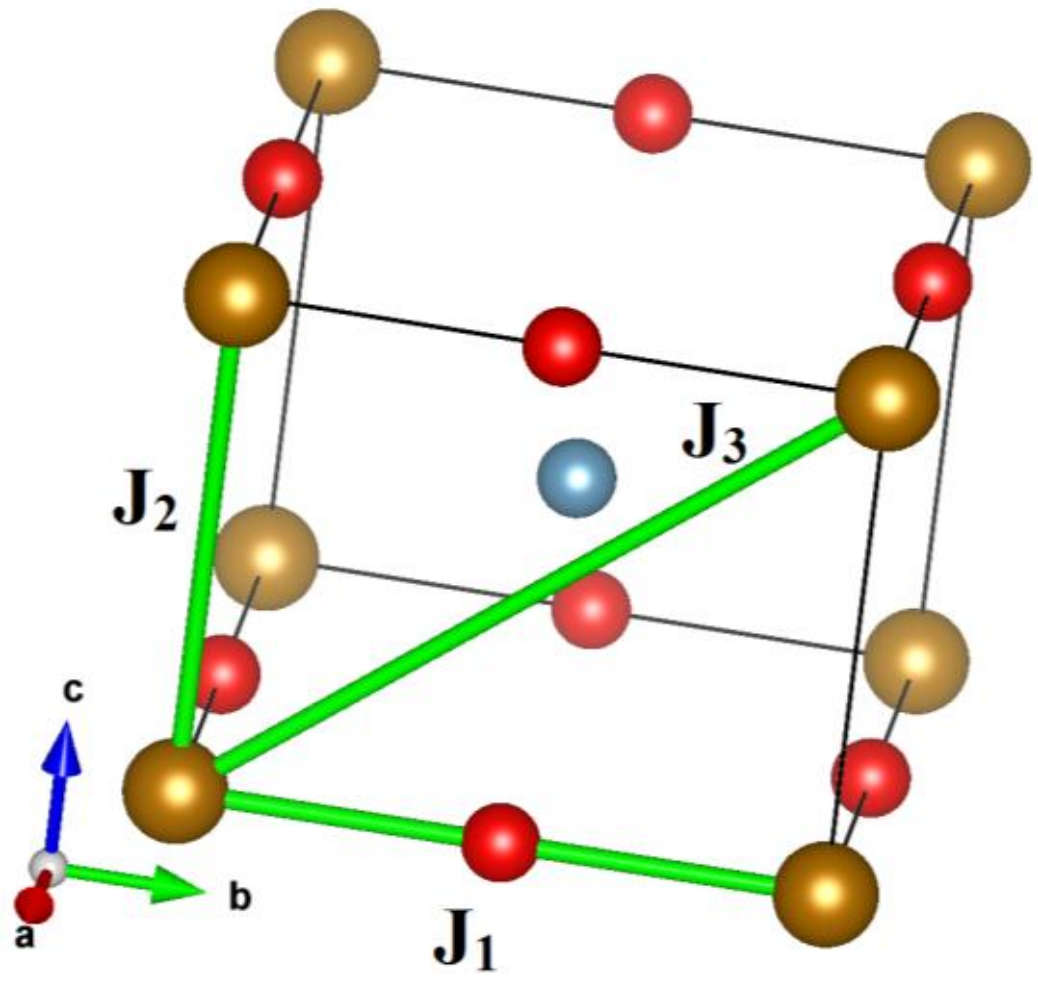

FIG. 12 (b) (Color online) The calculated magnetic exchange interaction parameters (*J*'s) in $SrFeO_2$ (*P*4/*mmm*) and p-$CaFeO_2$(*P*4/*mmm*) compound at different amplitude of $B_u$ phonon mode distortion.

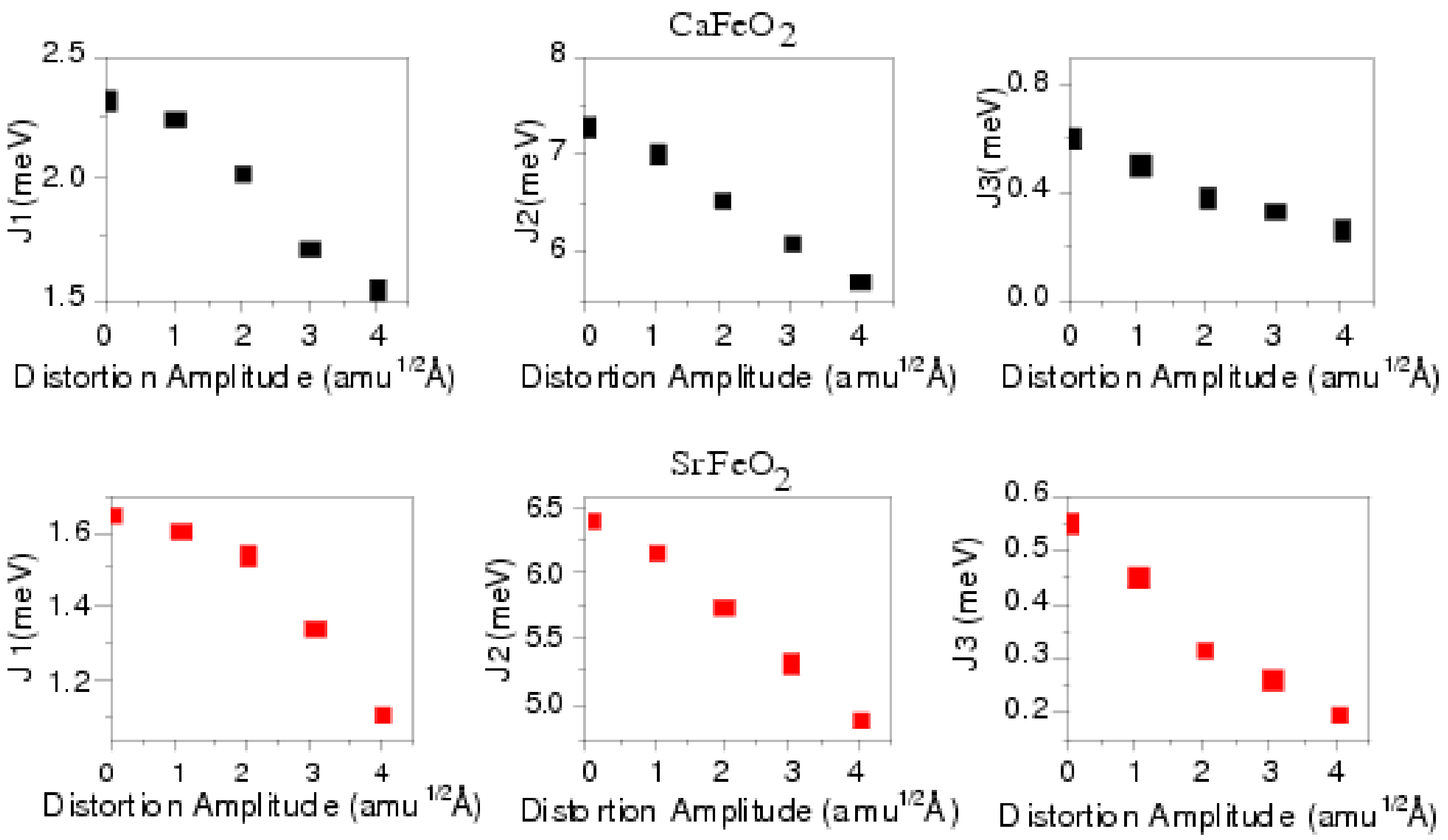

FIG. 13. (Color online) Calculated partial densities of states of various atoms in $SrFeO_2$ (*P*4/*mmm*) and $SrFeO_3$ (*Pm*-3*m*).

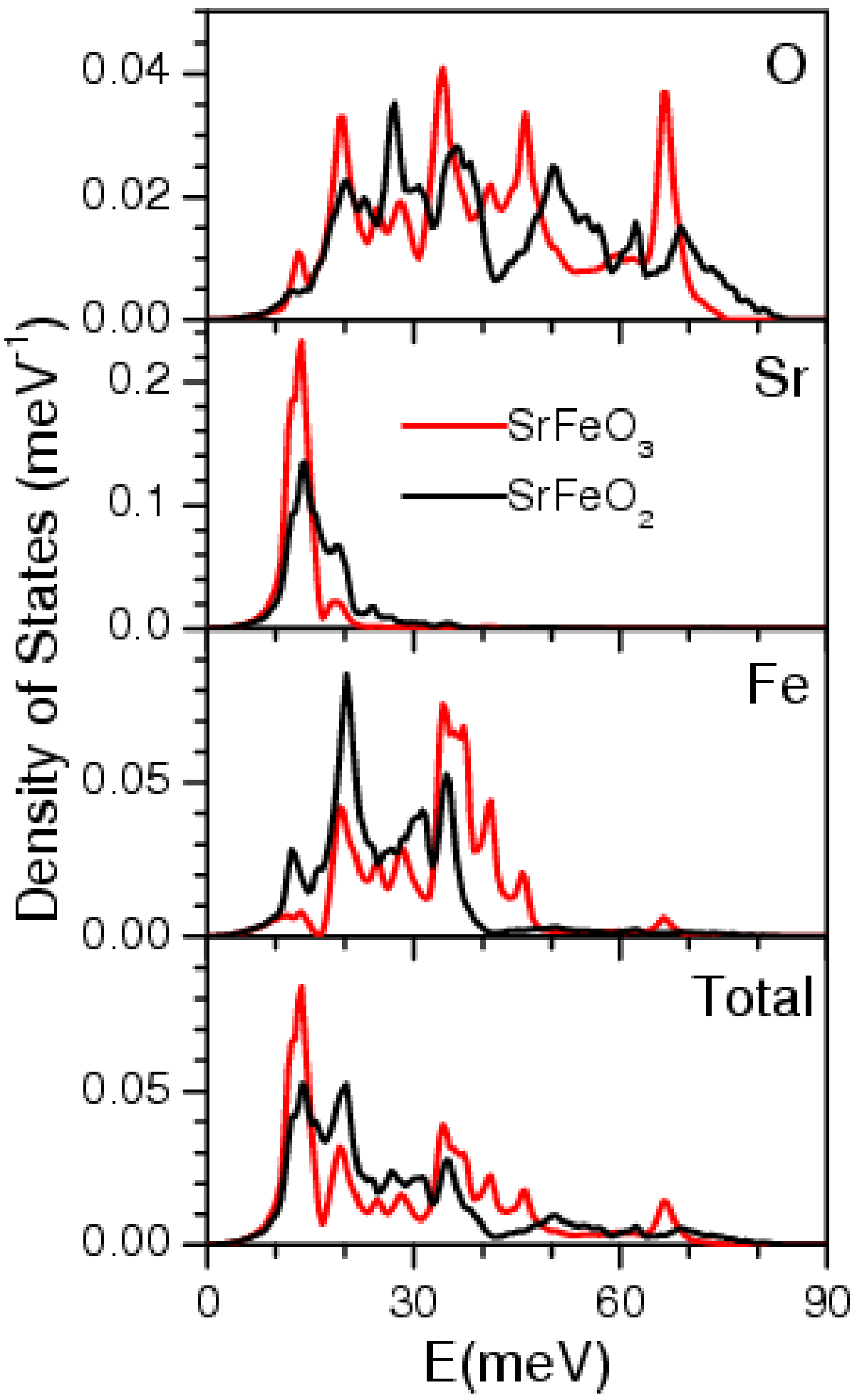